\documentclass[aps,twocolumn,prl,showpacs,preprintnumbers,amsmath,amssymb,superscriptaddress]{revtex4}

\usepackage{graphicx,epsfig}
\usepackage{dcolumn}
\usepackage{bm}
\usepackage{color}

\begin{document}

\title{Survival of superconducting correlations across the 2D
superconductor-insulator transition:  A finite frequency study}

\author{R. Crane}
\affiliation{Department of Physics and Astronomy, University of
California, Los Angeles, CA 90095}

\author{N.P. Armitage }
\thanks{npa@pha.jhu.edu}

\affiliation{Department of Physics and Astronomy, University of
California, Los Angeles, CA 90095}

\affiliation{D\'{e}partement de Physique de la Mati\`{e}re
Condens\'{e}e, Universit\'{e} de Gen\`{e}ve, quai Ernest-Ansermet
24, CH1211 Gen\`{e}ve 4, Switzerland}

\affiliation{Department of Physics and Astronomy, The Johns
Hopkins University, Baltimore, MD 21218 }

\author{A. Johansson}
\affiliation{Department of Condensed Matter Physics, Weizmann
Institute of Science, Rehovot 76100, Israel}

\author{G. Sambandamurthy}\affiliation{Department of Condensed
Matter Physics, Weizmann Institute of Science, Rehovot 76100,
Israel} \affiliation{SUNY - Buffalo, Department of Physics, 239
Fronczak Hall, Buffalo, NY 14260-1500}

\author{D. Shahar}
\affiliation{Department of Condensed Matter Physics, Weizmann
Institute of Science, Rehovot 76100, Israel}

\author{G. Gr\"{u}ner}
\affiliation{Department of Physics and Astronomy, University of
California, Los Angeles, CA 90095}

\date{\today}

\begin{abstract}

The complex AC conductivity of thin highly disordered InO$_x$
films was studied as a function of magnetic field through the
nominal 2D superconductor-insulator transition.  We have resolved
a significant finite frequency superfluid stiffness well into the
insulating regime, giving direct evidence for quantum
superconducting fluctuations around an insulating ground state and
a state of matter with localized Cooper pairs.  A phase diagram is
established that includes the superconducting state, a transition
to a `Bose' insulator and an eventual crossover to a `Fermi'
insulating state at high fields.  We speculate on the consequences
of these observations, their impact on our understanding of the
insulating state, and its relevance as a prototype for other
insulating states of matter that derive from superconductors.

\end{abstract}




\maketitle

\section{Introduction}

A quantum phase transition (QPT) is a zero temperature change of
state as a function of some non-thermal parameter (pressure,
doping, magnetic field, etc.) \cite{Sondhi,SachdevBook}.  The 2D
superconductor-insulator quantum phase transition (SIT) is a
particularly beautiful and illustrative one, reflecting a
transition between the two eigenstates at the extremes of a
superconductor's fundamental uncertainty relation between phase
and particle number ($ \Delta \theta \Delta n> 1$). One of the key
questions here concerns the nature of the zero temperature
destruction of the superconducting state. Does it proceed in a
mean-field fashion by destruction of the amplitude $\Delta$ of the
superconducting order parameter $ \psi = \Delta e^{i \theta}$ or
is it dominated by fluctuations of the superconducting phase
$\theta$ and by extension what is the nature of the insulating
ground state? The manner in which the superconductivity is
destroyed is an issue that has direct relevance to many other
important problems including that of high-temperature
superconductivity \cite{EmeryandKivelson,Ong,Corson}.

A related issue is the nature of the insulating state.  There have
been a number of proposals for an insulator with strong
superconducting correlations. For instance, Fisher and co-workers
\cite{FisherPRL,FisherPRB} postulated a dual description - the
so-called `dirty boson' model - of the SIT in which the
superconducting state reflects the condensation of Cooper pairs
and localization of vortices, while the insulating state is
characterized by condensed vortices and localized Cooper pairs.
More recently, it has been proposed that the transition is an
inhomogeneous one \cite{Ghosal}, and where global
superconductivity may obtain by percolation of locally
superconducting clusters \cite{Shimshoni}.
 To date, evidence for superconducting signatures in the
insulating state may have been found in an interesting positive
magnetoresistance in insulating films which has been interpreted
as a sign of vortex activation \cite{MarkovicVortex} as well as in
a crossover of the Hall coefficient $R_{xy}$ \cite{Paalanen} at a
field higher than the superconductor-insulator critical field
H$_{SIT}$. Although these results have been suggestive there has
been a notable lack of direct evidence for superconducting
correlations deep into the insulator. Moreover a number of
outlying issues remain. For instance, the Cooper pair gap appears
to close on the approach to the SIT in amorphous films, evidence
seemingly incompatible with the existence of localized Cooper
pairs \cite{TunnelValles1,TunnelValles2}. Still others have
challenged the existence of a direct transition between
superconducting and insulating states altogether and instead
postulate the existence of an intervening metal \cite{MasonMetal}.

In this paper we present the results of a comprehensive study of
the complex AC conductivity through the nominal 2D SIT in InO$_x$
thin films using microwave cavities.  This is the first such AC
study of this system;  all previous studies have used DC probes.
 Here, we have resolved a significant finite frequency superfluid
stiffness in the insulating state which persists well into the
strongly insulating regime and at fields up to 3 times the critical
field, giving direct evidence for a state with localized Cooper
pairs. This establishes a phase diagram with distinct regions
dominated by superconducting, `Bose' insulating, and `Fermi'
insulating effects.

\begin{figure}[h]
\centerline{\epsfig{figure=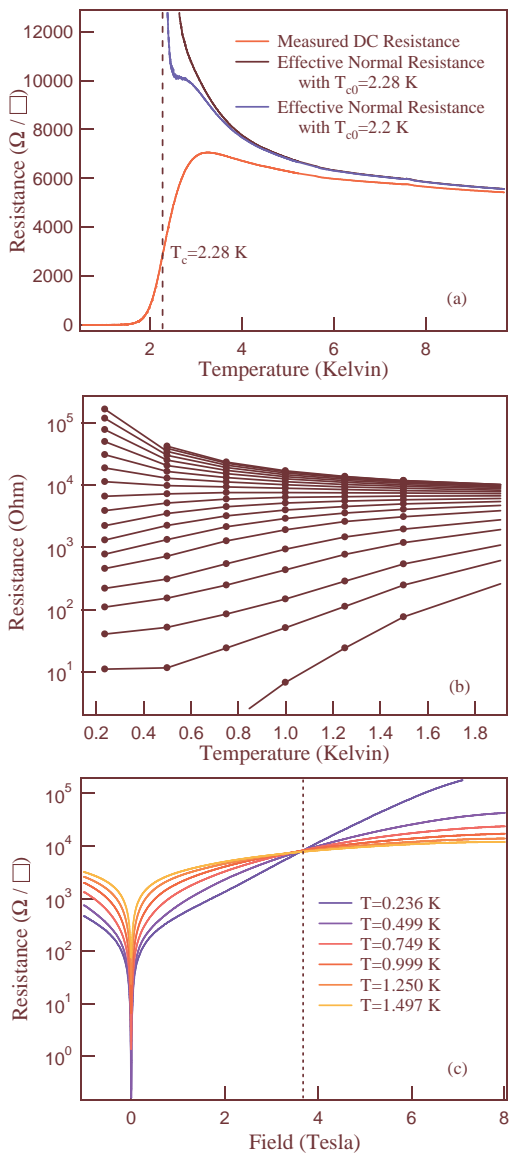,angle=0,width=7.0cm}}
\caption{(color) (a) Temperature dependence showing the broad
resistive transition resulting from fluctuations. The normal state
resistance curves are generated using the procedure described in
the text. The amplitude temperature T$_{co}$ is believed to be
2.28 K. (b) DC Sheet resistance vs. temperature for field values
H=0, 0.05, 0.1, 0.25, 0.5, 1, 1.5, 2.5, 3, 3.5, 4, 4.5, 5, 5.5, 6,
6.5, 7, 7.5, 8 Tesla. (c) Sheet resistance vs. magnetic field for
temperatures shown. The crossing point determines a critical field
of 3.68 Tesla. }
\end{figure}

\section{Experimental Details}

Samples were 200 $\AA$-thick 3mm-diameter highly-disordered
amorphous indium oxide ($\alpha$:InO) thin films prepared by e-gun
evaporating high purity (99.999\%) In$_2$O$_3$ on clean
19mm-diameter sapphire discs in high vacuum \cite{Kowal,Murthy04}.
The deposition methods are essentially the same as in the work of
Ref. \cite{Kowal} where it was demonstrated that amorphous InO$_x$
can be made reproducibly by e-beam evaporation of In$_2$O$_3$ and
then subsequent low-temperature annealing. In this way the end
result is quite different than films prepared via other methods
that create a granular or nanocrystallite morphology
\cite{HebardNakahara,Zhu}. Essentially identical films have been
used in a large number of recent studies of the 2D
superconductor-insulator quantum phase transition
\cite{Murthy04,Steiner1,Steiner2,Murthy05,Murthy06,CranePRB,Gantmakher}.
Post-deposition, samples were room-temperature annealed for
approximately one week in ambient air and then, except for short
periods, were kept well below LN$_2$ temperatures for the duration
of the measurements (approximately 2 months).

Thin machined aluminum masks were used to pattern the films
creating a 200 $\AA$-thick 3mm-wide circular amorphous film
centered on the disc.  Deposition was well controlled and samples
with specific properties can be made reproducibly
\cite{Murthy04,Steiner2}.  For structural characterization, we
deposit two more films along with the sample: (1) onto a TEM grid
for electron diffraction. (2) for AFM scans. We believe the films
to be morphologically homogeneous with no crystalline inclusions
or large scale morphological disorder for a number of reasons. The
TEM-diffraction patterns are found to be diffuse rings with no
diffraction spots, suggesting a completely amorphous layer with no
crystalline inclusions. Moreover, the AFM images show continuous
films with no voids or cracks and are in fact entirely featureless
down to the limiting resolution of the AFM (a few nanometers). We
also note that the R vs T curves when investigating the 2D
superconductor-insulator transition \cite{Murthy04} in these films
are smooth with no re-entrant behavior that has been the hallmark
of gross inhomogeneity.

Experiments were performed in a novel cryomagnetic resonant
microwave cavity system.  The cavity diameter was chosen to
optimize performance in the 22 GHz ($\hbar \omega / k_B = 1.06$ K)
TE011 mode, but measurements were possible at a number of discrete
frequencies from 9 to 106 GHz. Cavities were operated in a low
power regime where the cavity response was independent of the
input intensity to ensure no sample heating.

Various well-known relations were used to relate the resonances'
frequency shift $\Delta \omega$ and change in inverse quality
factor $\Delta \frac{1}{Q}$ upon sample introduction to the
complex conductivity
\cite{KotzlerBrandt1,KotzlerBrandt3,Redbook,Peligrad}.  Extensive
details of the data analysis scheme can be found elsewhere
\cite{CranePRB}.  We only mention here that since the films are
grown in the center of a sapphire disc and are hence centered
along the cavity's axis, we can make use of both in-plane electric
fields or perpendicular magnetic fields. The symmetry of the
sample's placement means that, for all TE modes if there is an
in-plane electric field at the sample position, the out-of-plane
field is zero and vice versa. Different analysis schemes are used
for the two possibilities. Again, see Ref. \cite{CranePRB} for a
very thorough discussion of our precise experimental arrangement
and analysis scheme.  A consequence of the sample's extreme
thinness is that the real part of the conductivity is essentially
proportional to the change in inverse quality factor $\Delta
\frac{1}{Q}$ and the imaginary part of the conductivity is
proportional to frequency shift and there is only a weak mixing
between real and imaginary components.  We of course used the
complete analysis, which can be found in Ref. \cite{CranePRB}. The
conversions from $\Delta \frac{1}{Q}$ and $\Delta \omega$ to
conductivity were made by normalizing the AC resistance of the
cavity data to DC data at temperatures well above the occurrence
of superconductivity and insisting that the superfluid stiffness
(defined below), was frequency \textit{independent} at zero field
as $T \rightarrow 0$.

DC resistance was measured on co-deposited samples in a two-probe
configuration by low frequency AC lock-in techniques using
excitation currents of $10 pA-10nA$. The probe's lead resistances,
which have a negligible temperature dependence in the displayed
temperature range, were well characterized and have been
subtracted from the displayed data.

\section{Results}

The zero field DC resistance curve shown in Fig. 1a is fairly
typical for a highly disordered superconducting thin film showing
an approximately log increase of $R$ with decreasing temperature,
before the occurrence of a broad superconducting transition.  As
shown in Fig. 1b, when the field is increased the resistance
curves cross over from a superconducting positive $d R/d T$ to an
insulating negative $d R/d T$ behavior similar to previous
field-tuned studies \cite{Paalanenscaling,Yazdani,Goldmanscaling}.
We emphasize the very strong temperature dependence of the
insulating state.  At low temperatures and 8 $T$ the InO$_x$ films
are very strongly insulating with a resistance that is approaching
$10^6 \Omega/\square$.  When plotted as $R$ vs $H$ (Fig. 1c) the
experimental data shows a low-temperature iso-resistance crossing
point of 3.68 Tesla, which can be identified as the critical field
of the SIT, $H_{SIT}$. The DC data can be shown to be consistent
with previous studies
\cite{Goldmanscaling,Yazdani,Paalanenscaling} that found scaling
as a function of the reduced variable $|H-H_{SIT}|/T^{1/z \nu}$
where $z \nu$ is consistent with the exponent for 2D classical
percolation $\frac{4}{3}$.

As the temperature is lowered, distinct regions typified by
amplitude and then phase fluctuations of the superconducting order
parameter $ \psi = \Delta e^{i \theta}$ are expected.  Using the
procedure of Gantmakher \cite{Gantmakher,CranePRB}, a lower bound
on the temperature where the amplitude becomes well defined
$T_{co}$ can be estimated as the lowest temperature (Fig. 1a) that
does not cause an inflection point in the extracted effective
normal state resistance $R_N(T)$ as defined by the full expression
for the Aslamazov-Larkin fluctuation resistivity \cite{AL}.  In
two dimensions, this extracted temperature scale $T_{c0} = 2.28 K$
does not signify the occurrence of a phase transition, but instead
represents the temperature scale below which the Cooper pair
amplitude becomes well defined.

We now turn our attention to the AC conductivity.  At low
temperatures, the imaginary conductivity for a long-range ordered
superconductor is expected to have the form $\sigma_{2} = \frac{N
e^2}{\omega m}$ where $N$ is the superfluid density and $e$ and
$m$ are the electronic charge and mass respectively.  For a
fluctuating superconductor one can define $ \sigma_2 =
\frac{N(\omega) e^2}{\omega m}$  where an additional frequency
dependence is captured by a generalized frequency dependent
superfluid density. The superfluid density is directly
proportional to the superfluid stiffness which is the energy scale
for inducing slips in the superconducting phase. The use of a
frequency dependent density or stiffness has been the usual
treatment within, for instance, the finite frequency treatment of
the Kosterlitz-Thouless-Berezinskii (KTB) transition
\cite{KTdynamics}. In Fig. 2a we display the $H=0$ generalized
frequency dependent superfluid stiffness, $T_{\theta}$ (in degrees
Kelvin) extracted via the relation $\sigma_2 = \sigma_Q \frac{k_B
T_{\theta}}{\hbar \omega}$, where $\sigma_Q = \frac{4e^2}{hd} $ is
the quantum of conductance for Cooper pairs divided by the film
thickness.  We emphasize that the superfluid stiffness
$T_{\theta}$ is $not$ a temperature, but is an energy scale
expressed in temperature units.  The superfluid stiffness curves
all cross the predictive line ($ T_{\theta} = 4 T_{KTB}$) for the
KTB transition in a range near 1.6 K, however a frequency
dependence is acquired by the superfluid stiffness at a lower
temperature of approximately 1.1 K.  We associate this lower
temperature as the KTB scale below which the superconductor
becomes robust against vortex phase fluctuations. A somewhat
similar phenomenology has been seen in quasi 2D high temperature
superconductors at THz frequencies \cite{Corson}.  The $T_{KTB}$
of $\approx 1.1$ K and $ T_{c0}$ of 2.28 K give the temperatures
at zero field for phase coherence and a well defined
superconducting amplitude, respectively.  A detailed study of
these fluctuation regimes at H=0 can be found elsewhere
\cite{CranePRB}.

\begin{figure}[h]
\centerline{\epsfig{figure=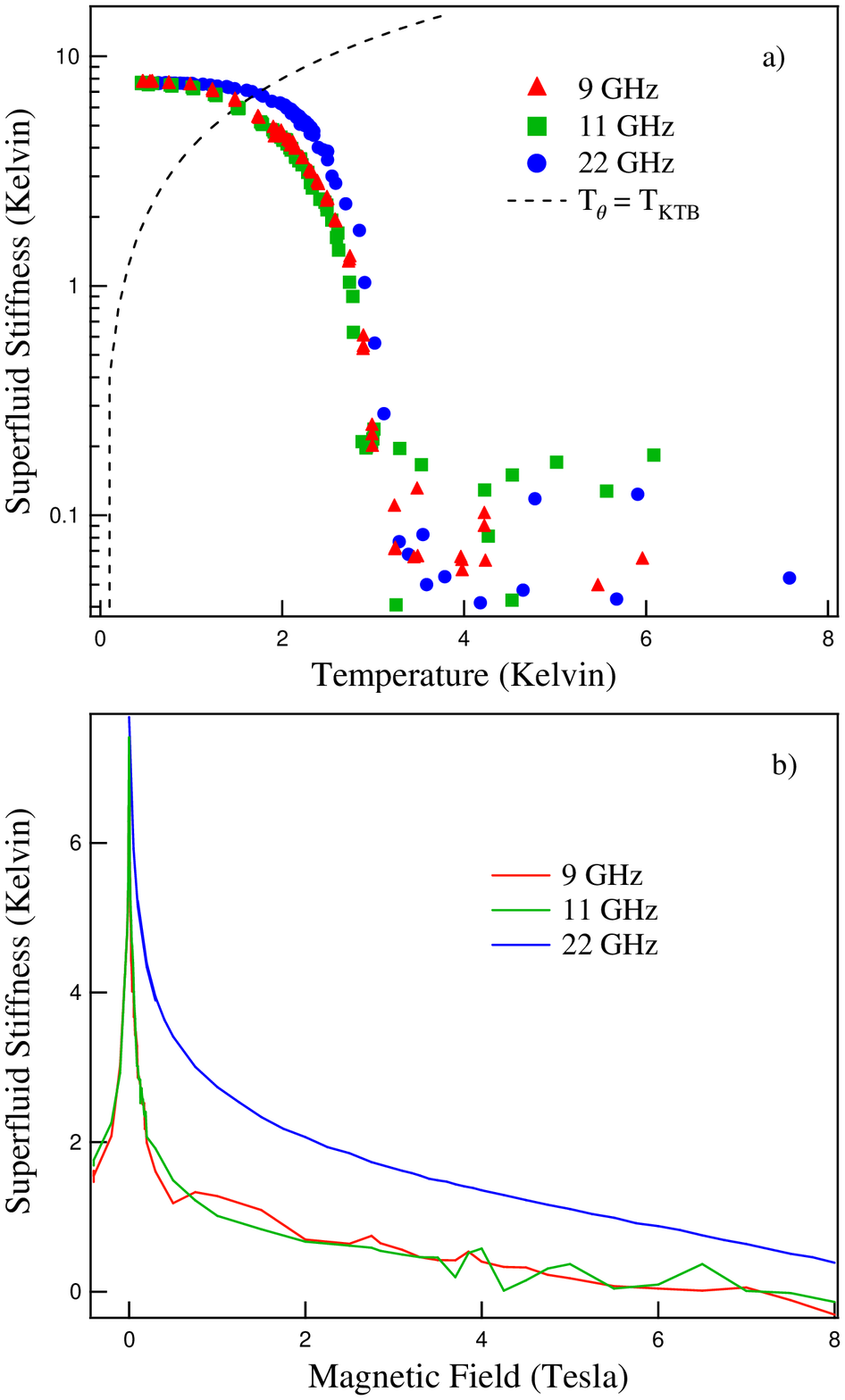, angle=0,width=7.0cm}}
\caption{(color) Superfluid stiffness, $T_{\theta}$ ($\propto
\omega \sigma{_2}$), extracted as described in the text. (a)
Temperature dependence at H=0.  Also shown is the theoretical
$T_{KTB}$ which intersects $T_{\theta}$ around $T = 1.1 K$. (b) T=
0.5 K magnetic field dependence at select low frequencies.}
\end{figure}

Using the BCS relation $2 \Delta / k_B T_{c0} = 3.53$ and the
measured $T_{c0} = 2.28$ K, all of our displayed operating
frequencies are well below the BCS expectation of 168 GHz for
above threshold gap excitation at low temperature and zero field.
 In principle however, there may be a normal electron contribution to
the response from both thermally excited electrons as well as from
above gap excitations when $\hbar \omega > 2 \Delta$ on the
approach to T$_c$ or at high fields where the gap's size may
decrease.  In practice however, these give a negligible
contribution to the $\sigma_2$ response in a highly disordered
material like ours as the motion of these essentially normal
electronic excitations are highly damped.  We can give a rough
estimate of this contribution by a very approximate calculation as
follows.

Mangetoresistance measurements on very similar samples have given
a coherence length of approximately 6 nm \cite{Steiner2}, which is
set by the normal state electron mean free path in highly
disordered materials.  This gives the very large effective
scattering rate ($1/\tau$) of approximately 200 - 300 THz (using a
rough estimate for the Fermi velocity (c/200)), which is
reasonable based on the high disorder and large normal state
resistivity of our sample. An estimate based on the normal state
resistivity, using a free electron mass and a carrier density
estimated from Hall measurements on similar samples
($10^{21}/$cm$^3$) \cite{ShaharOvadyahu} gives a scattering length
of the same order of a magnitude.

Using this scattering rate, our definition of $T_{\theta}$ and its
measured low temperature value (7.66 K), and the Drude relations
we can estimate the maximum possible $contamination$ contribution
to the measured stiffness from ``normal" electrons as $T_{\theta
,cont} = T_{\theta}[T\rightarrow0] \frac{(\omega \tau)^2}{1 +
(\omega \tau)^2 }$.  This is the contribution if all the spectral
weight in the superconducting delta function was scattered in a
normal state electron-like fashion.  At even our highest
frequency, this contribution of $2 \times 10^{-6}$ K is far
smaller than the experimental sensitivity.  Even if the effective
scattering rate was overestimated by even a factor of 100 the
contamination contribution would only rise to approximately 0.02 K
which is still well less than our sensitivity. Hence, the
contribution from ``normal'' electrons is a completely
insignificant contribution to $\sigma_2$ and hence $T_{\theta}$
and we can safely assume that all of our $\sigma_2$ signal is due
to a superconducting contribution of some variety.

Displayed in Fig. 3 is the finite-frequency complex response at $H
\neq 0$ for the 22 GHz data.  Similar plots can be made at other
frequencies.  Fig. 3a shows the real part of the AC resistance
measured over a range of temperatures and magnetic fields.  The
bold line is a contour defined by the AC resistance at the
critical field (defined by the DC measurements) extrapolated to
finite temperatures.  We identify the region below this critical
contour of $R = 2295 \Omega $ (in bold) as the domain over which
the underlying $T=0$ superconducting phase has influence.

In Fig. 3b is shown the superfluid stiffness $T_{\theta}$ at 22
GHz, using the same conversion as above, and plotted as a function
of temperature and magnetic field.  We observe that the superfluid
stiffness is finite well into the insulating regime and even at
fields and temperatures where the DC resistance can be inferred to
be over 10$^6\Omega /\Box$.  The use of relatively high probing
frequencies allows us to resolve superconducting fluctuations into
the insulating part of the phase diagram.  An additional advantage
of our high frequencies is that they are at least two orders of
magnitude higher than a generous estimate for the vortex depinning
frequency \cite{Henderson} found in thin films of conventional
superconductors.  The vortex contribution to the optical response
is therefore expected to be purely dissipative, leaving the
superfluid as the only principal contributor to $\sigma_2$
\cite{mixing}.  With a low frequency probe one would have the
additional complication of being sensitive to the vortex
polarization as they are displaced from the pinning sites which
would give an additional contribution to $\sigma_2$. The
finite-frequency superfluid stiffness falls quickly with
increasing field, but remains finite above $H_{SIT}$, well into
the insulating regime to fields almost 3 times the critical field
H$_{SIT}$. This is the first direct measure of superconducting
correlations well into the insulating side of the 2D
superconductor-insulator transition in an amorphous film.

\begin{figure}[h]
\centerline{\epsfig{figure=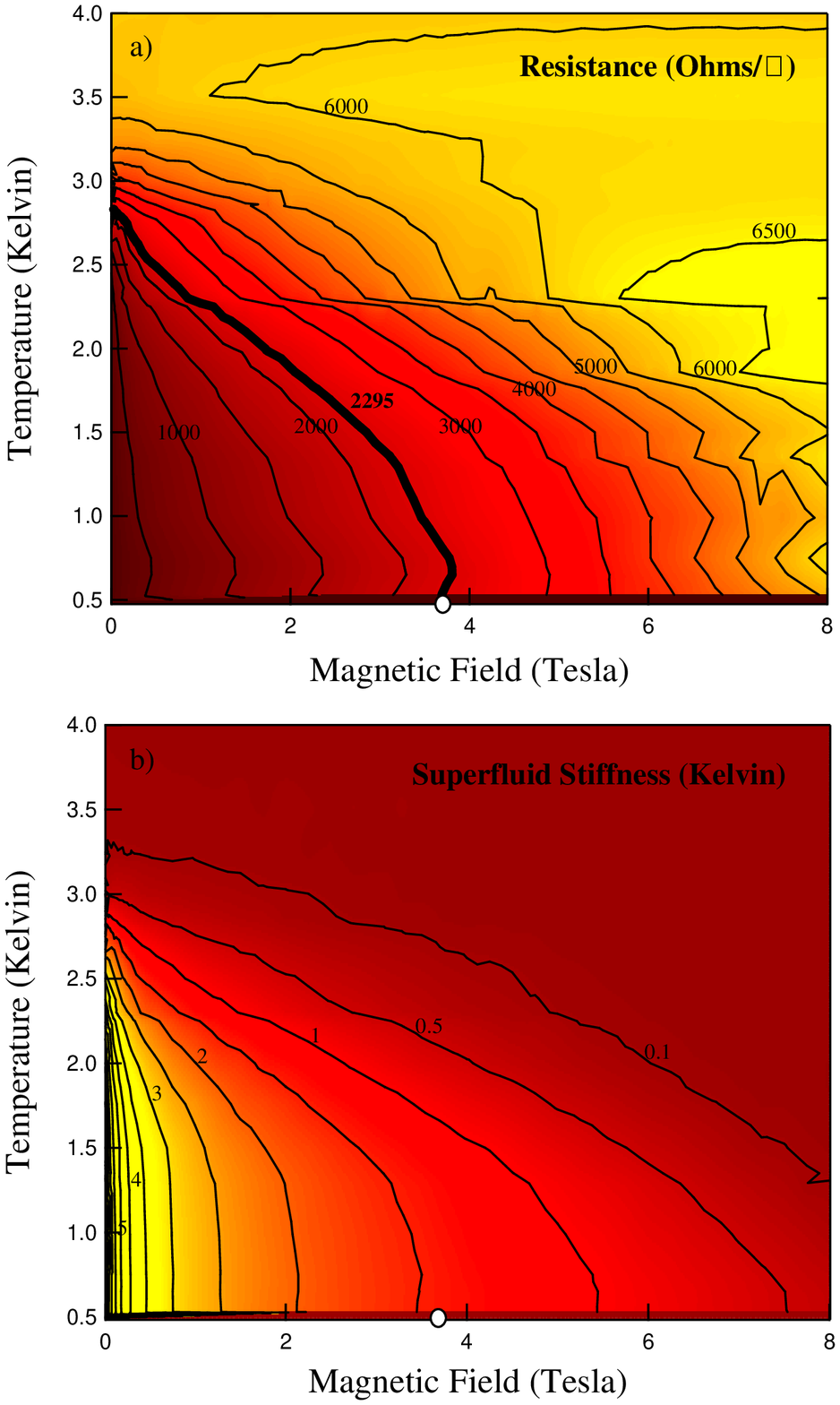,clip=,angle=0,width=7.0cm}}
\caption{(color) (a) AC Sheet resistance ($\propto
Re(\frac{1}{\sigma d})$).  The bold line is the critical
resistance extrapolated to finite temperature.  (b) Superfluid
stiffness ($\propto \omega \sigma_2$) at 22 GHz. The critical
field, $H_{SIT}$, is shown as a white dot at $H=3.68 T$. Yellow
indicates maximum.}
\end{figure}

\section{Discussion}

Our observation of a finite frequency superfluid stiffness at
$H>H_{SIT}$ is not inconsistent with an insulating T=0 ground
state. As alluded to above, our experiments are sensitive to
superfluid \textit{fluctuations} because we probe the system on
short time scales via an experimental frequency $\omega_{exp}$
that is presumably high compared to an intrinsic order parameter
fluctuation rate $\omega_{QC}$ close to the transition. Above
$H_{SIT}$, an experimental probe in the limit $\omega_{exp} \ll
\omega_{QC}$ will not detect superfluid fluctuations as a system
with $T_\theta(\omega \rightarrow 0) \neq 0$ can support
superfluid flow, behavior obviously incompatible with the notion
of an insulator.  On general grounds an insulator with
$\sigma_1=0$ at zero frequency, must have a $\sigma_2$ that is
negative at $\omega = 0^+$ by Kramers-Kronig considerations,
meaning that such a system can not appear superconducting at low
frequencies. It is only by using a relatively high $\omega$ probe
we detect these fluctuations.  Returning to Fig. 2b we see the
dramatic drop in superfluid stiffness with increasing field and
the strong frequency dependence of $T_{\theta}$ which reflects
that the Cooper pairs are correlated on time scales that can be
resolved at the probing frequency, so that while long range order
does not exist at finite T, short-range correlations do.  We also
note that at our lowest temperatures and well into the insulating
side of the phase diagram, that although the magnitude of the
superfluid stiffness is of the same order as it is above $T_{c0}$,
the signal becomes temperature independent at low temperature
showing the intrinsic quantum mechanical nature of these
fluctuations.

\begin{figure}[h]
\centerline{\epsfig{figure=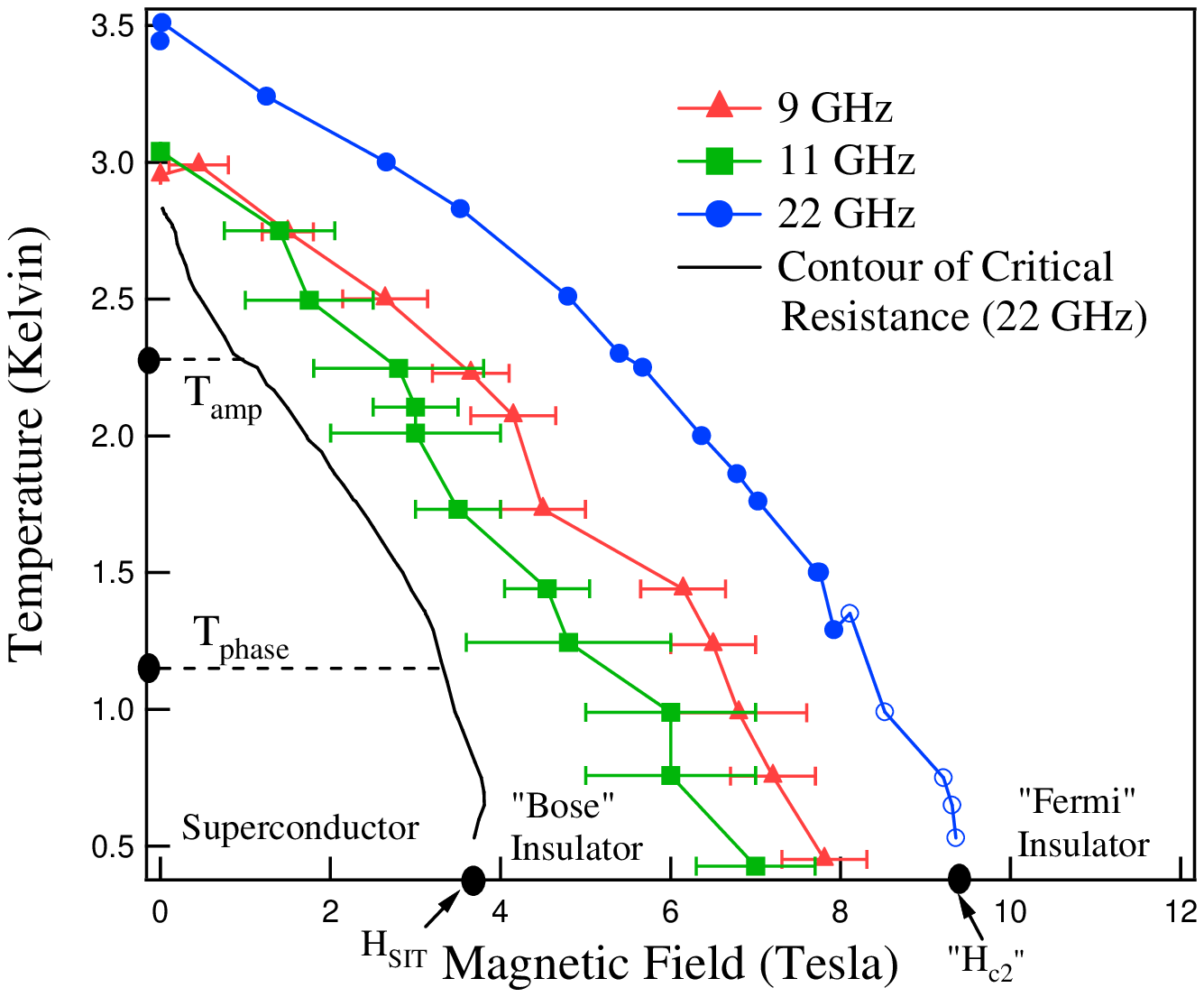,clip=,angle=0,width=8.0cm}}
\caption{(color) 2D Field Tuned Superconductor-Insulator ``Phase
Diagram''. Contours with markers are the detection limit for
$T_{\theta}$ at specified frequencies. Black dots on the temperature
axis denote $T_{amp}$ ($=T_{c0}$) and $T_{phase}$ ($=T_{KTB}$),
signifying temperatures where the amplitude and phase respectively
become well-defined. $H_{SIT}$ appears as a black dot and the
contour of critical resistance from Fig. 3a appears as a solid black
line. Open symbols represent data obtained from a small linear
extrapolation beyond our maximum field of 8 Tesla.}
\end{figure}

We are able to extract a phase diagram that establishes the
existence of superconducting correlations well into the insulating
state. In Fig. 4 contours are plotted which denote the region
above which our superfluid stiffness becomes almost
indistinguishable from the normal state noise level (set at 1 \%
of the $T \rightarrow 0$, $H=0$ superfluid stiffness), thereby
giving a measure of the extent of superconducting correlations
into the insulating regime.  The noise is greatest for frequency
contours away from our cavity's optimal operating frequency of 22
GHz, but it is evident that the higher frequency probe at 22 GHz
allows one to examine the fluctuations of the order parameter
persisting at fields higher than $H_{SIT}$ and temperatures higher
than $T_{c0}$, at shorter length and time scales than the 9 or 11
GHz probes. In the high-frequency limit one would expect that the
detection limit would eventually extrapolate to a field where the
Cooper pairs are completely depaired.  In the low frequency limit
we would expect that the superfluid density would disappear
at $H_{SIT}$.  For 22 GHz we observe that the data extrapolates to
a value of H=9.35 T, which is close to the value of the pair
breaking field, $H_{c2}$, found in similar films of InO$_x$
\cite{Murthy04,Steiner1,Steiner2} giving evidence that 22 GHz is
near the high-frequency limit.  It is impossible to say with our
current results whether this pair breaking scale should be
associated with orbital or spin effects. We do note that this
field is well above the naive estimate ($H_P = 1.86 \times T_c$)
of the Pauli pair breaking field, but that disorder is known to
lead to an enhanced $H_{c2}$ over the expectation from its H=0
value \cite{Maekawa}.  Our results and interpretation are not
consistent with the inference of Ref.\cite{Steiner1} that Cooper
pairs still had integrity at fields many times the pair breaking
scale.  It would interesting to verify whether the crossover
observed in $R_{xy}$ by Paalanen \textit{et al.} \cite{Paalanen} -
which has also been interpreted as a `Bose' glass/`Fermi' glass
crossover - is found in our films at the extrapolated pair
breaking field scale.  Additionally it would be important to
measure the Nernst effect in InO$_x$ films.  Such measurements
have recently revealed a finite Nernst signal at fields above
$H_{SIT}$ in NbSi that has been interpreted in terms of
superconducting fluctuations \cite{Behnia}.

\section{Conclusions}

We have performed the first AC conductivity study across the 2D
superconductor-insulator quantum phase transition.  We find
evidence for a finite superfluid stiffness at fields well into the
insulating phase giving direct evidence for superconducting
fluctuations around an insulating ground state.  We emphasize that
we believe the only true phase transition in the system
(neglecting for a moment the possibility of an intervening metal
phase) is at H$_{SIT}$ and that the contours in Fig. 4 give a
frequency dependent crossover.  We observe superconducting
correlations not just asymptotically close to the SIT as one might
for localized electrons, but an extended region above $H_{SIT}$.
This establishes for the first time a model-free, unambiguous
picture of an insulating state that is dominated by
superconducting correlations - a `Bose' insulator. We conclude
that at some level this `Bose' insulator is best described by
localized Cooper pairs in contrast to localized electrons.  Our
phase diagram is characterized by a superconducting ground state,
a phase transition to a `Bose' insulating state at H$_{SIT}$ and a
crossover to `Fermi' insulator near a depairing field
``H$_{c2}$''.

This work raises certain questions about the expected
electrodynamic response of the various proposed phases of matter.
 It might be expected, for instance, that the functional dependence of
the DC $R(T)$ is different for `Fermi' vs `Bose' insulators, but
these differences may be subtle \cite{DasDoniach} and difficult to
distinguish experimentally.  In the present case, we have a
heuristic argument about the detection of finite frequency
superfluid stiffness as being indicative of the `Bose' insulator,
but detailed calculations are lacking.  We encourage theorists to
perform these calculations as, in principle, the relevant response
function is well-defined and calculable as the finite frequency
generalization of the usual zero frequency superfluid stiffness.
 Moreover, it has been argued that an anomalous metallic state
intervenes between the superconducting and insulating states at
low temperature \cite{MasonMetal}.  At finite temperature the
difference in the dissipative response between a fluctuating
superconductor and this metal could be subtle, but the difference
in the reactive response may be dramatic. Again, detailed
calculations are lacking.  In either case we have shown that the
underlying superconductivity will manifestly have to be taken into
account to describe the insulating and anomalous metallic states.

Finally, it has not escaped our notice that this work has a direct
connection to many current theoretical proposals within the
context of high-temperature superconductivity.  Our results show
in principle that such insulating ground states characterized by
substantial quantum superconducting fluctuations can exist and
serve as a prototype for other insulating states of matter that
derive from superconductors \cite{granath,zlatko,chen,podolsky}.

\section{ Acknowledgements }

The authors would like to thank  E.H. Brandt, S. Chakravarty, J.
Deisenhofer, C. Grimaldi, A. Kapitulnik, S. Kivelson, N.
Markovi\'{c}, D. Peligrad, Z. Te\v{s}anovi\'{c}, and E. van Heumen
for helpful discussions and commentary on the manuscript. The
authors would also like to thank V. Zaretskey for help with the
instrumentation development and K. Holczer for experimental
support at a crucial time in this project. Research at UCLA was
supported by the NSF (DMR-0454540). Research at the Weizmann
Institute was supported by the ISF, the Koshland Fund and the
Minerva Foundation.  NPA has also been supported by the NSF's
International Research Fellowship Program and MaNEP.


\begin{thebibliography}{99}

\bibitem{Sondhi} S. L. Sondhi \textit{et al.}, Rev. Mod.
Phys. \textbf{69}, 315 (1997).

\bibitem{SachdevBook} S. Sachdev, \textit{Quantum Phase
Transitions}, Cambridge University Press (1999).

\bibitem{EmeryandKivelson} V.J. Emery, S.A. Kivelson, Nature \textbf{374}, 434 (1995).

\bibitem{Ong} Z. A. Xu \textit{et al.}, Nature
\textbf{406}, 486 (2000).

\bibitem{Corson} J. Corson \textit{et al.}, Nature \textbf{398}, 221 (1999).

\bibitem{FisherPRL} M. P. A. Fisher, Phys. Rev. Lett. \textbf{65}, 923 (1990).

\bibitem{FisherPRB} M. C. Cha \textit{ et al.}, Phys. Rev. B \textbf{44}, 6883 (1991).

\bibitem{Ghosal}  Amit Ghosal, Mohit Randeria, and Nandini
Trivedi, Phys. Rev. B \textbf{81}, 3940 (1998).

\bibitem{Shimshoni}  E. Shimshoni, A. Auerbach, A. Kapitulnik, Phys. Rev. Lett. \textbf{80}, 3352 (1998).

\bibitem{MarkovicVortex} N. Markovic \textit{et al.}, Phys. Rev. Lett. \textbf{81}, 701 (1998).

\bibitem{Paalanen} M. A. Paalanen, A. F. Hebard, R. R. Ruel, Phys. Rev. Lett. \textbf{69}, 1604 (1992).

\bibitem{TunnelValles1}  J. M. Valles, Jr., R. C. Dynes, and J. P. Garno, Phys. Rev. Lett. 69, 3567–3570
(1992).

\bibitem{TunnelValles2} S.-Y. Hsu, J.A. Chervenak, J.M. Valles, Phys. Rev. Lett. \textbf{75}, 132 (1995).

\bibitem{MasonMetal} N. Mason, A. Kapitulnik, Phys. Rev. Lett. \textbf{82}, 5341 (1999).

\bibitem{Kowal} D. Kowal and Z. Ovadyahu, Sol. St. Comm. \textbf{90}, 783
(1994).

\bibitem{Murthy04} G. Sambandamurthy \textit{et al.}, Phys. Rev. Lett. \textbf{92}, 107005 (2004).

\bibitem{HebardNakahara} A. F. Hebard and S. Nakahara, Appl. Phys. Lett. \textbf{41}, 1132
(1982).

\bibitem{Zhu}  F. Zhu \textit{et al.}, Thin Solid Films 359, 244
(2000).

\bibitem{Steiner1} M.A. Steiner, G. Boebinger, and A. Kapitulnik, Phys. Rev. Lett. \textbf{94}, 107008
(2005).

\bibitem{Steiner2} M.A. Steiner,  A. Kapitulnik, Physica C \textbf{422},
16 (2005).

\bibitem{Murthy05} G. Sambandamurthy et al.,  Phys. Rev. Lett. \textbf{94}, 017003
(2005).

\bibitem{Murthy06} G. Sambandamurthy et al., Europhysics Lett. \textbf{75}, 611
(2006).

\bibitem{CranePRB} R. Crane \textit{et al.}, Phys. Rev. B
\textbf{65}, 094506 (2007).

\bibitem{Gantmakher} V. F. Gantmakher and M.V. Golubkov, JETP Letters
\textbf{73}, 131 (2001).

\bibitem{KotzlerBrandt1} E. H. Brandt, Phys. Rev. Lett. \textbf{71}, 2821
(1993).

\bibitem{KotzlerBrandt3} J. K\"{o}tzler \textit{et al.}, Phys. Rev. B \textbf{50}, 3384 (1994).

\bibitem{Redbook} O. Klein, S. Donovan, M. Dressel, G.
Gr\"{u}ner, \textit{Int. J. Infrared Millim. Waves} \textbf{14},
2423 (1993).

\bibitem{Peligrad} D.-N. Peligrad \textit{ et al.}, Phys. Rev. B \textbf{64}, 224504
(2001).

\bibitem{Paalanenscaling} A.F. Hebard and  M.A. Paalanen, Phys. Rev. Lett. \textbf{65}, 927
(1990).

\bibitem {Yazdani} A. Yazdani and  A. Kapitulnik, Phys. Rev. Lett. \textbf{74}, 3037 (1995).

\bibitem {Goldmanscaling} Y. Liu \textit{ et al.}, Phys. Rev. Lett. \textbf{67},
2068 (1991).

\bibitem{AL} L. G. Aslamazov and A. I. Larkin, Phys. Lett. \textbf{26A}, 238 (1968).

\bibitem{KTdynamics} V. Ambegaokar \textit{et al.}, Phys. Rev. B \textbf{21},
1806–1826 (1980).

\bibitem{ShaharOvadyahu} D. Shahar and Z. Ovadyahu, Phys. Rev. B \textbf{46}, 10917 (1992).

\bibitem{Henderson} W. Henderson \textit{ et al.}, Phys. Rev. Lett. \textbf{80}, 381
(1998).

\bibitem{mixing}  In contrast to the usual convention in vortex physics \cite{vortexref}
we parameterize out-of-phase current response as $\sigma_2$, and
not $\frac{1}{\omega L}$, as the later quantity will diverge in a
system with residual dissipation and vanishing superfluid density.
Because vortex conductivities add in \textit{parallel} with the
charge response, a weak mixing into $\sigma_2$ of the in- and
out-of-phase vortex response is possible which means that our
quantity T$_\theta$ needs to be rigorously interpreted as a
$lower$ $bound$ on the superfluid energy. However, a simple
estimate based on our extracted vortex viscosity shows that this
contribution can not be more than a few percent at our highest
fields.

\bibitem{vortexref} M. W. Coffey, J. R. Clem, Phys. Rev. Lett. \textbf{67}, 386
(1991).

\bibitem{Maekawa} S. Maekawa, H. Ebisawa, and H. Fukuyama, J. Phys. Soc. Jpn. \textbf{52}, 1352 (1983).

\bibitem{Behnia} A. Pourret \textit{et al.}, Nature Physics
\textbf{2}, 683 (2006).

\bibitem{DasDoniach} D. Das and S. Doniach, Phys. Rev. B \textbf{57}, 14440 - 14443
(1998).

\bibitem{granath} M. Granath \textit{et al.}, Phys. Rev. Lett. 87 167011 (2001).

\bibitem{zlatko} Z. Te\v{s}anovi\'{c}, Phys. Rev. Lett. 93, 217004 (2004).

\bibitem{chen} H.-D. Chen \textit{et al.}, Phys. Rev. Lett.
\textbf{93}, 187002 (2004).

\bibitem{podolsky} D. Podolsky \textit{et al.}, Phys. Rev. B \textbf{67},
094514 (2003).


\end{thebibliography}

\end{document}